# Monolayer Semiconductor Auger Detector

Authors: Colin M. Chow[1], Hongyi Yu[2,3], John R. Schaibley[1,4], Pasqual Rivera[1], Joseph Finney[1], Jiaqiang Yan[5], David G. Mandrus[5,6], Takashi Taniguchi[7], Kenji Watanabe[7], Wang Yao[3], David H. Cobden[1]*, Xiaodong Xu[1,8]*

[1]Department of Physics, University of Washington, Seattle, Washington 98195, USA.
[2]Guangdong Provincial Key Laboratory of Quantum Metrology and Sensing & School of Physics and Astronomy, Sun Yat-Sen University (Zhuhai Campus), Zhuhai 519082, China.
[3]Department of Physics and Centre of Theoretical and Computational Physics, University of Hong Kong, Hong Kong, China.
[4]Department of Physics, University of Arizona, Tucson, Arizona 85721, USA.
[5]Materials Science and Technology Division, Oak Ridge National Laboratory, Oak Ridge, Tennessee 37831, USA.
[6]Department of Materials Science and Engineering, University of Tennessee, Knoxville, Tennessee 37996, USA.
[7]National Institute for Materials Science, Tsukuba, Ibaraki 305-0044, Japan.
[8]Department of Materials Science and Engineering, University of Washington, Seattle, Washington 98195, USA
*email: cobden@uw.edu, xuxd@uw.edu

## Abstract

**Auger recombination in semiconductors is a many-body phenomenon in which recombination of electrons and holes is accompanied by excitation of other charge carriers. Being nonradiative, it is detrimental to light emission. The excess energy of the excited carriers is normally rapidly converted to heat, making Auger processes difficult to probe directly. Here, we employ a technique in which the Auger-excited carriers are detected by their ability to tunnel out of the semiconductor through a thin barrier, generating a current. We employ vertical van der Waals (vdW) heterostructures with monolayer $WSe_2$ as the semiconductor and the wide band gap hexagonal boron nitride (hBN) as the tunnel barrier to preferentially transmit high-energy Auger-excited carriers to a graphite electrode. The unambiguous signatures of Auger processes are a rise in the photocurrent when excitons are created by resonant excitation, and negative differential photoconductance resulting from the shifts of the exciton resonances with voltage. We detect holes Auger-excited by both neutral and charged excitons, and find that the Auger scattering is surprisingly strong under weak excitation. The selective extraction of Auger carriers at low, controlled carrier densities that is enabled by vdW heterostructures illustrates an important addition to the techniques available for probing relaxation processes in 2D materials.**

## Introduction

The two-dimensional (2D) monolayer semiconductors of formula $MX_2$ (M = Mo, W; X = S, Se) have direct optical band gaps[1,2]. Nevertheless, when pristine exfoliated monolayers are photoexcited, nonradiative recombination usually dominates[3,4]. This is in part a consequence of Auger processes[5], whose rates are enhanced relative to those in 3D semiconductors because of stronger Coulomb interactions[6]. Among these processes is exciton-exciton annihilation, which dominates at high excitation density[7–9], and results in either an excited electron in the conduction band or a hole in the valence band. At lower densities, however, opinions vary on the significance of different Auger contributions. One possibility is thermal activation of trapped photocarriers and subsequent increase in the number of delocalized carriers conducive to Auger scattering[10]. This process has been argued to become more important at elevated temperatures[11,12], although the lack of temperature dependence in some measurements argues against the significance of such processes[13,14]. In addition, in the tungsten-based materials there are dark exciton ground states which may provide a phonon-assisted Auger channel at low excitation powers[15], though the significance of this too is unclear[16]. The ambiguous situation is a consequence of the fact that, despite their ubiquity, Auger processes are hard to probe because they are both ultrafast and nonradiative.

In this work, we employ an unconventional photocurrent technique to reveal exciton/trion-hole Auger scattering in monolayer $WSe_2$. From the dependence of the photocurrent on electrode voltage and excitation energy, we can extract spectral information and band offsets. Figure 1a is a schematic and 1b an optical micrograph of the device which we will focus on in the main text. We obtained consistent results with other similar devices (Supplementary Information (SI) §6). A $WSe_2$ monolayer flake is sandwiched between thin hexagonal boron nitride (hBN) dielectric layers, in this case with thicknesses of 8 nm (top) and 10 nm (bottom). Few-layer graphene (FLG) is used for electrical contacts CT1 and CT2 to the $WSe_2$ monolayer. Their separation of about 5 μm defines the $WSe_2$ channel length. Another FLG piece on top serves both as a gate for electrostatically doping the $WSe_2$ and as an optically transparent electrode for collecting carriers that cross the hBN barrier. A further pair of split FLG bottom gates, labeled BG1 and BG2, is included for doping the $WSe_2$ at the respective contacts to reduce the contact resistance. (See Methods and SI §1 for fabrication details).

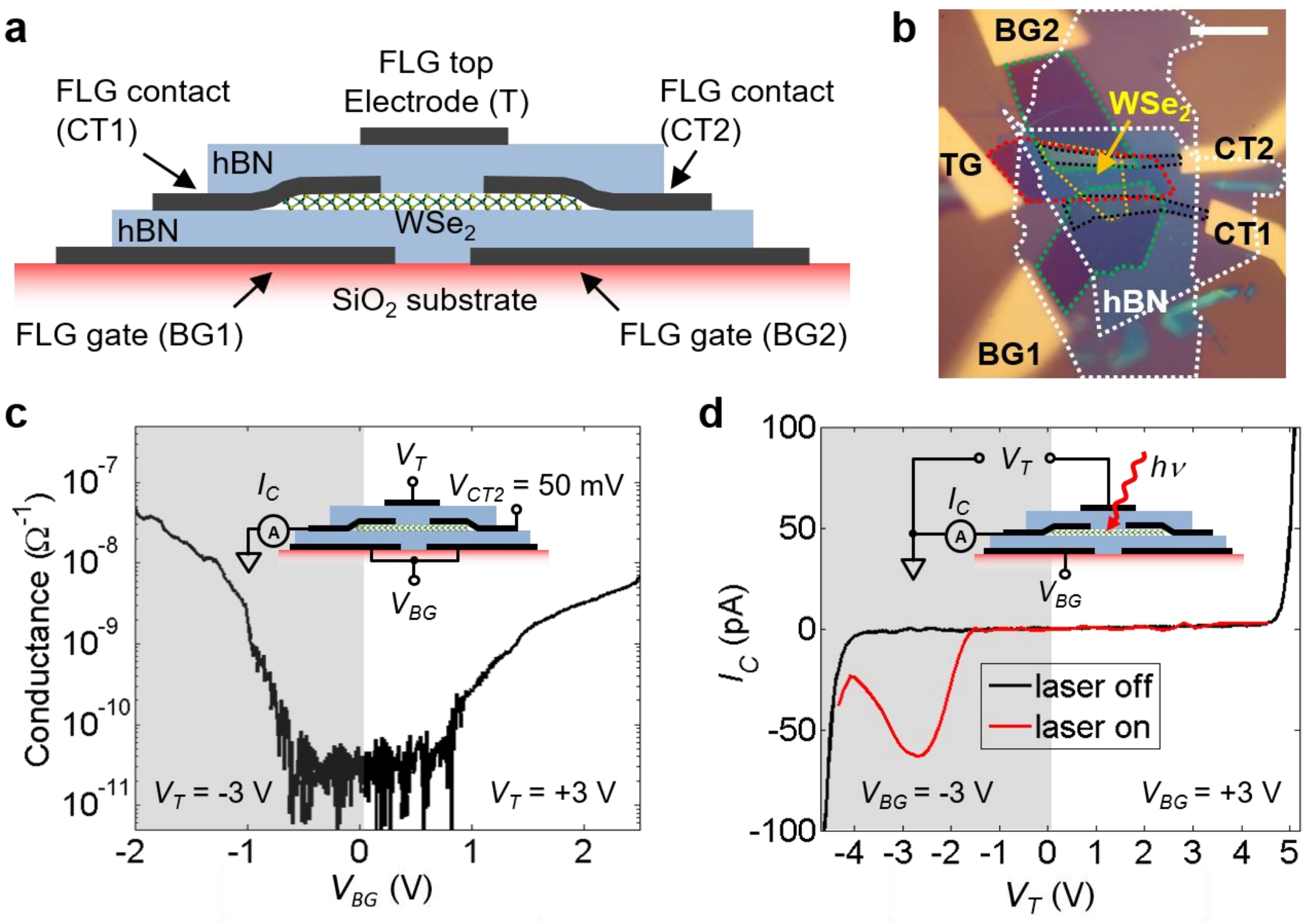

**Fig. 1 | Device geometry and basic characterization. a,** Schematic of a device with monolayer $WSe_2$ as the active layer. Metal electrodes connecting the few-layer graphene (FLG) flakes are omitted for clarity. **b,** Optical micrograph of device 1, measurements on which are presented in the main text. Dotted lines indicate the boundaries of individual flakes, whose thicknesses can be found in SI §1. Scale bar: 15 μm. **c,** Semi-log plot of the conductance between CT2 and CT1, as a function of the bottom gate bias $V_{BG}$, measured at a temperature of 5 K. In the shaded (non-shaded) region, $V_T$ = -3 (+3) V. Inset: measurement configuration. **d,** Gate current ($I_C$) between CT1 and top electrode (T) versus $V_T$ with (red) and without (black) laser illumination at $\hbar\omega$ = 1.71 eV (1 μW at 725 nm). Here CT2 and BG2 are disconnected, as indicated in the inset. In the shaded (non-shaded) region, $V_{BG}$ = -3 (+3) V.

# Results

### Device operating condition in Auger photocurrent mode

Fig. 1c shows the conductance between CT1 and CT2 as a function of back-gate voltage $V_{BG}$, applied equally to BG1 and BG2, measured in the dark with a bias of 50 mV on CT2. We obtain ambipolar operation at low temperature (here 5 K), implying suitably conducting contacts for either electrons or holes, by setting the top electrode voltage $V_T$ to -3 V when $V_{BG} < 0$ and to +3 V when $V_{BG} > 0$. This dopes the channel with the same carrier type as the contact regions (see SI §2). In a similar manner, whenever we vary $V_T$ we set $V_{BG}$ to either +3 V or -3 V as appropriate to keep the contacts conducting.

This device structure permits multiple photocurrent spectroscopy modes (see SI §3). In Auger photocurrent mode, we measure the current $I_C$ that flows from the top electrode through the thin hBN to contact CT1, keeping CT2 disconnected. Similar measurement results were obtained using contact CT2 instead. The $I_C$-$V_T$ characteristic in the dark (black trace in Fig. 1d) is typical for an hBN tunneling barrier of thickness 8 nm with low defect density[17]: the current is negligible at biases smaller than about 4 V and rises rapidly at larger biases due to Fowler-Nordheim tunneling. To have negligible dark current, and to avoid degrading the hBN, we keep the magnitude of $V_T$ smaller than 4 V in the following measurements.

### Exciton- and trion-induced Auger photocurrent

When a laser of frequency $\omega$ is focused to a spot (~1 μm in diameter) between the contacts, appreciable photocurrent can be generated, depending on $V_T$ and $\hbar\omega$. For example, at $\hbar\omega$ = 1.71 eV (red trace in Fig. 1d), photocurrent appears when $V_T$ is more negative than $-2$ V, rises to a peak at $V_T \approx -2.7$ V, and then exhibits negative differential photoconductance[18] (NDPC), decreasing to a minimum at $-4$ V. The dependence of $I_C$ jointly on $V_T$ and $\hbar\omega$ in the hole-doped regime ($V_{BG} = -3$ V and $V_T$ ranging negative) is shown as an intensity plot in Fig. 2a. A corresponding plot of the optical absorption (see SI §4) is shown in Fig. 2b, where the peaks due to the neutral A and B excitons ($X_A^0$ and $X_B^0$) and positive trion ($X_A^+$) of monolayer $WSe_2$ are labeled. Evidently, $I_C$ shows features associated with these absorption resonances (labeled accordingly). In the cases of $X_A^0$ and $X_A^+$, $I_C$ exhibits a peak as a function both of $\hbar\omega$ and of $V_T$. Fig. 2c shows traces of $I_C$ versus $V_T$ for selected photon energies close to the trion absorption resonance. Multiple peaks can be seen, each with an associated region of NDPC.

In Fig. 2d we plot the positions of the peaks in both $I_C$ (blue) and absorption (red) for $X_A^0$ and $X_A^+$ as a function of $\hbar\omega$ and $V_T$, derived from the data in the red boxes in Figs. 2a and 2b respectively. The absorption peaks blue-shift substantially with increasingly negative $V_T$. This implies a reduction of the exciton/trion binding energy that exceeds the bandgap renormalization, as seen previously[19]. The observation that the $X_B^0$ peak does not blue-shift with $V_T$ is consistent with this understanding, because the initial electron state for $X_B^0$ generation is in the lower spin-split valence band far below the Fermi level. The close correspondence between the features in the photocurrent and the absorption strongly implies that the photocurrent is related to the rate of exciton generation. The NDPC occurs when the absorption resonances blue-shift above the excitation energy as $V_T$ increases, thereby reducing the exciton generation rate.

We also plot in Fig. 2d, in black, the positions of the $X_A^0$ and $X_A^+$ photoluminescence (PL) peaks measured in the same device. At small $V_T$ they match the absorption peak positions, but as $V_T$ increases the PL peaks red-shift and so diverge from the blue-shifting absorption peaks. This can be understood as a consequence of the fact that PL is not sensitive to the occupancy of valence band states, combined with free-carrier screening that renormalizes the band gap downwards[20].

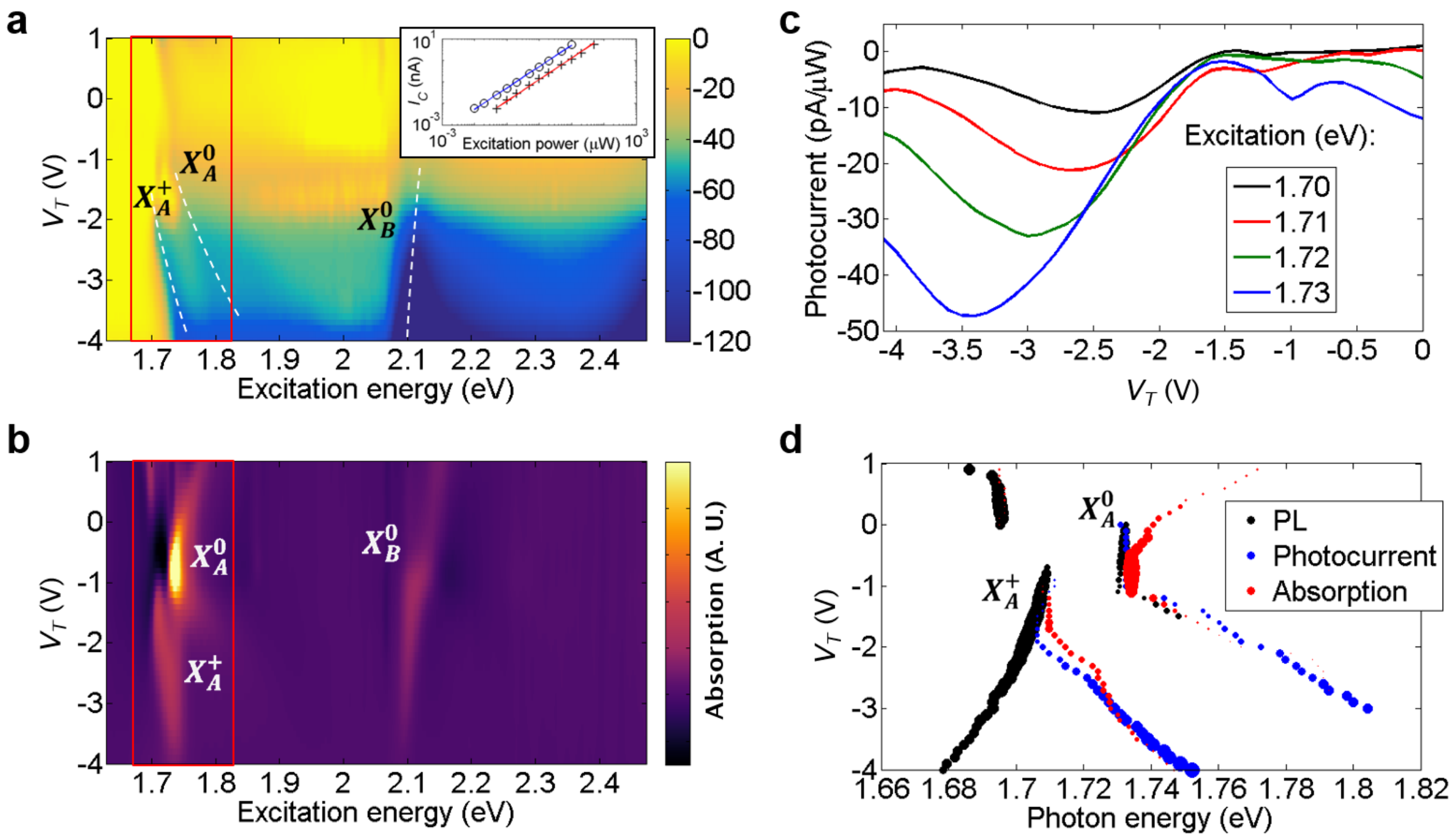

**Fig. 2 | Photocurrent and absorption spectra in the hole-doped regime. a,** Intensity plot of photocurrent $I_C$ as a function of $V_T$ and excitation energy (at $V_{BG} = -3$ V). White dashed lines indicate positions of the exciton/trion peaks. The A exciton ($X_A^0$), A trion ($X_A^+$), and B exciton ($X_B^0$) features are labeled. Colour bar: pA/μW. Inset: peak photocurrent vs. excitation power, showing linear behavior, for resonant excitation of $X_A^+$ (crosses) at 1.725 eV and $X_B^0$ (circles) at 2.108 eV. **b,** Intensity plot of optical absorption strength measured simultaneously with the photocurrent. Colour bar: arbitrary linear scale. **c,** $I_C$ versus $V_T$ at selected photon energies near the $X_A^0$ and $X_A^+$ resonances. **d,** Comparison of peak positions extracted from the data in the red boxes in **a** and **b** and also from photoluminescence measurements (SI §2).

## Mechanism of Auger hole detection

The sign of photocurrent here implies that illumination causes holes to flow from the $WSe_2$ through the hBN to the top electrode. One possible mechanism for this is direct photoexcitation of holes to states in the $WSe_2$ valence bands near or below the hBN valence band edge, from which they can simply pass over the barrier. This is energetically possible since the photon energy is much larger than the $WSe_2$–hBN valence band offset, $E_{VBO}$~ 0.8 eV, deduced from recent measurements of $WSe_2$–graphene[21] and graphene–hBN[22] band offsets. However, direct one-photon absorption is parity forbidden[23], and moreover its rate should be independent of excitonic effects. Similar arguments also rule out direct photo-activation of mid-gap charged defects in hBN. Instead, the fact that the photocurrent has peaks near the exciton absorption resonances implies that it depends on the exciton population, and therefore that the passage of holes through the barrier is assisted by excitons.

In the simplest case, a hole is excited to below the hBN valence band edge by the Auger recombination of a single exciton. This process is energetically possible, since the exciton energy is much greater than $E_{VBO}$, and unlike for one-photon absorption the process is not parity forbidden and the rate should be proportional to the exciton population, which is greatest near the absorption resonances. The hole is injected far from the Fermi energy in the graphite, where, as in any metallic electrode, the quasi-particle lifetime is very short and in-plane momentum is not a good quantum number; hence momentum conservation places no constraint. Since $I_C$ is simply proportional to the laser power (see Fig. 2a inset) this must be the dominant process, because Auger processes involving more than one exciton would produce a superlinear power dependence. With this understanding, using rate equations (see SI §7) we can estimate a lower bound for the exciton-hole Auger rate of $10^{10}$ $s^{-1}$. This is surprisingly large compared with Auger rates in highly doped bulk semiconductors[24] of between $10^6$ and $10^8$ $s^{-1}$.

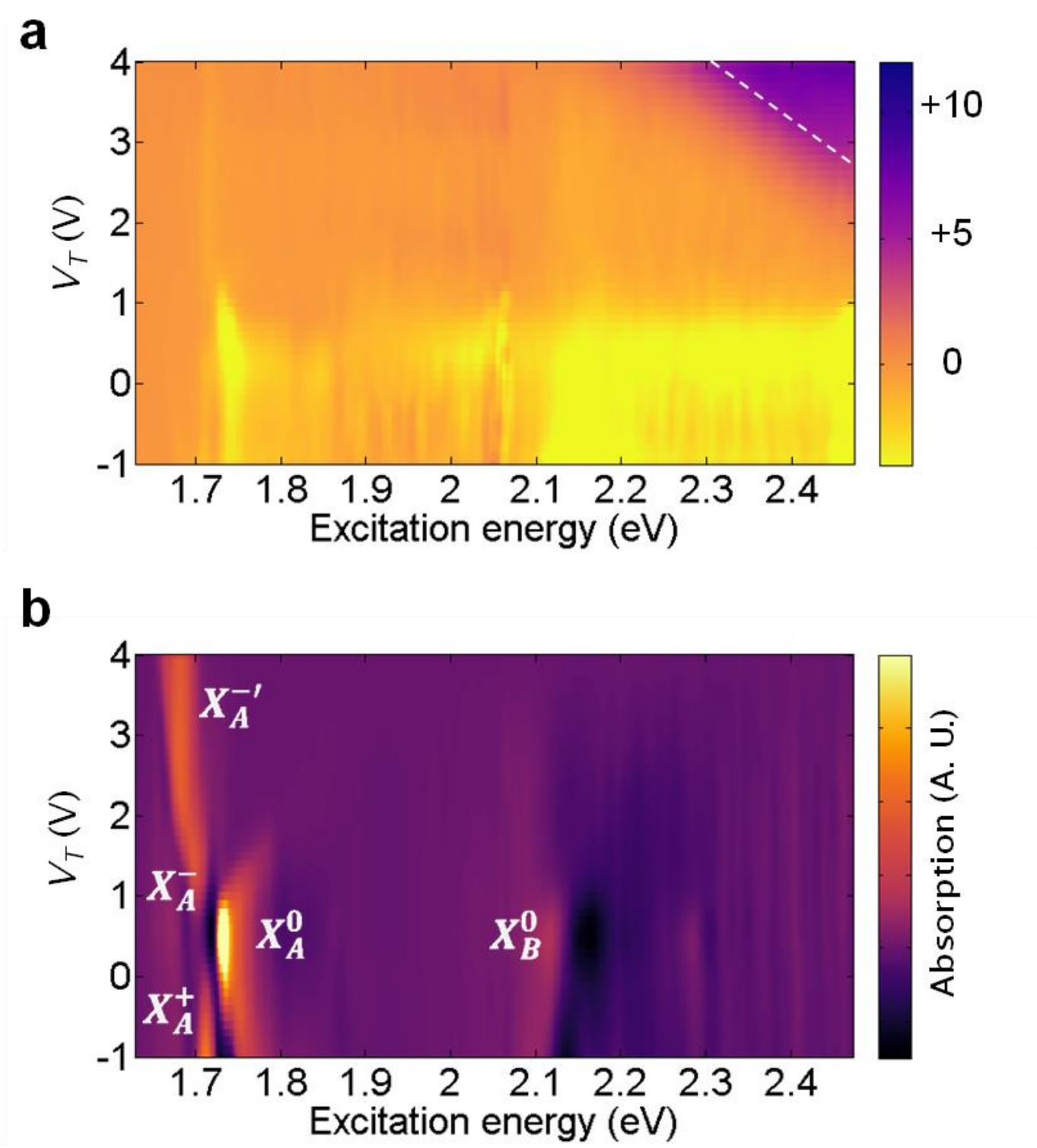

**Fig. 3 | Photocurrent and absorption spectra in the electron-doped regime. a,** Intensity plot of photocurrent $I_C$ as a function of $V_T$ and excitation energy (at $V_{BG} = +3$ V). Photocurrent is

seen only above a threshold indicated by the white dashed line. Colour bar: pA/μW. **b,** Intensity plot of concurrently measured absorption, with excitonic species labeled.

Figures 3a and b show measurements corresponding to those in Figs. 2a and b, respectively, but here in the electron-doped regime ($V_{BG} = +3$ V and $V_T$ ranging positive). Unlike in the hole-doped regime, no photocurrent at all is seen for $\hbar\omega$ < 2.2 eV, even though a negative trion ($X_A^-$) resonance is visible at $\hbar\omega \approx$1.7 eV in the absorption. This is explained by the $WSe_2$–hBN conduction band offset, $E_{CBO}$, being much larger than the energy a conduction-band electron can gain by either one-photon absorption or Auger recombination of $X_A^-$. Using $E_{VBO} = 0.8$ eV together with $WSe_2$ and hBN band gaps of 2.1 eV[25] and 6.0 eV[26], respectively, gives $E_{CBO}$~3.0 eV. Photocurrent does however flow at higher $\hbar\omega$, above a bias threshold which is indicated by the white dashed line in Fig. 3a. This can be explained by direct one-photon absorption by electrons at the $WSe_2$ conduction band edge[23], which is parity-allowed, immediately followed by tunneling though the hBN barrier, whose transparency increases with increasing electric field. The linear decrease of the bias threshold with $\hbar\omega$ can be reproduced well using the WKB approximation assuming a step height of 3.0 eV that matches $E_{CBO}$ (see SI §9).

# Discussion

The interpretation of the data discussed above is summarized in Fig. 4. A schematic $I_C$ versus $V_T$ traces is plotted in Fig. 4a, showing NDPC in the hole-doped regime and photo-assisted tunneling in the electron-doped regime. First, consider negative $V_T$, where the $WSe_2$ is hole-doped (Fig. 4b). When $\hbar\omega$ is resonant with $X_A^+$, trions are generated (left), and when an electron and a hole in a trion recombine the excess energy is transferred by an Auger process to the remaining hole, which then has enough energy to pass through the hBN valence band and produce photocurrent (right). Similarly, when $\hbar\omega$ is resonant with $X_A^0$ or $X_B^0$, one of the generated neutral excitons can excite a free hole when it recombines[24], or it can capture a free hole to form a trion and then recombine exciting the hole. Second, consider increasingly negative $V_T$ starting from one of these resonant conditions (Fig. 4c). This causes the absorption resonance to blue-shift above $\hbar\omega$, suppressing exciton generation and so reducing the photocurrent, resulting in NDPC. Third, consider positive $V_T$, where the $WSe_2$ is electron-doped (Fig. 4d). Because of the large $E_{CBO}$, neither one-photon absorption nor Auger processes can excite electrons to high enough energy to pass into the hBN conduction band. However, at large enough $V_T$ and $\hbar\omega$ (Fig. 4d) the barrier transparency for photo-excited electrons is sufficient for photo-assisted tunneling to occur and give photocurrent, though without NDPC.

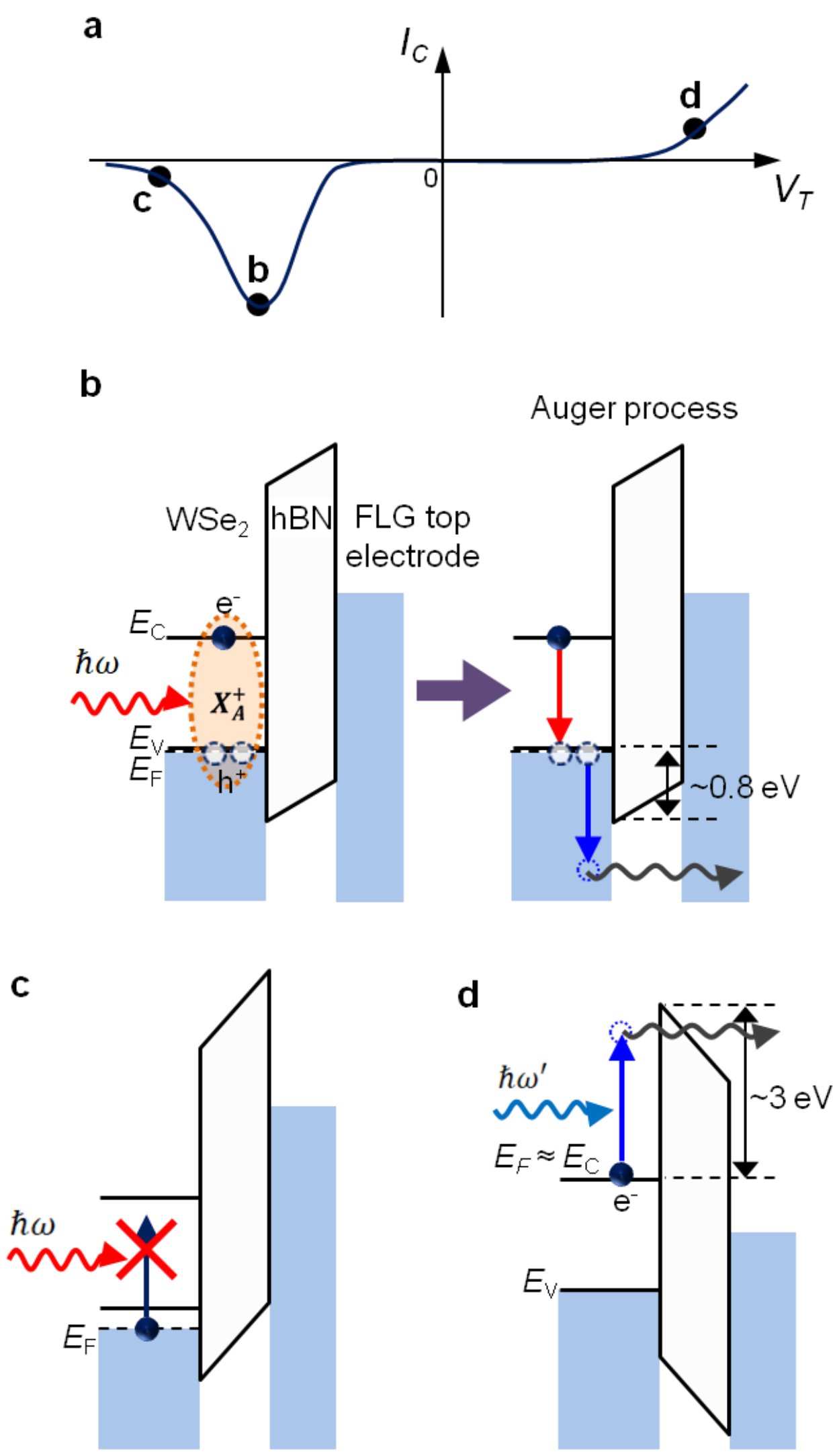

**Fig. 4 | Schematic summary of NDPC mechanism in hole-doped regime ($V_T$ < 0 V) and photo-assisted tunneling in electron-doped regime ($V_T$ > 0 V). a.** A schematic $I_C$ versus $V_T$ trace showing NDPC in hole-doped regime and photo-assisted tunneling in electron-doped regime **b.** When $V_T$ is negative the WSe$_2$ is p-doped. At a certain value of $V_T$ (point **b** in $I_C$ versus $V_T$ plot), $\omega$ is in the middle of the $X_A^+$ absorption resonance and $X_A^+$ trions are most rapidly created (left panel). Recombination of an electron and hole in a trion causes Auger scattering of the remaining hole to an energy below the hBN valence band edge, allowing it to pass through the hBN producing photocurrent (right panel). **c,** As $V_T$ is made more negative (point **c**), the $X_A^+$ resonance blue-shifts, causing the rate of creation of excitons and hence the photocurrent to decrease. **d,** When $V_T$ is positive the WSe$_2$ is n-doped. The large conduction band offset here prevents photoexcitation directly over the barrier, but at large enough $V_T$ (point **d**) photo-excited electrons can tunnel across the hBN barrier.

In conclusion, we observe excitation-frequency dependent photoconductance peaks that result from Auger processes linked to the excitonic absorption resonances in the monolayer semiconductor WSe$_2$. We find that the dominant Auger process in our WSe$_2$ heterostructures is excitation of holes by recombination of individual $X_A^0$ excitons or $X_A^+$ trions, and we infer a lower bound of $10^{10}$ $s^{-1}$ on the rate. While the mechanism of hot Auger carrier extraction resembles that considered in other systems[27–29], our method of using van der Waals heterostructures enables the study of Auger processes at low excitation density and with gate control of the doping. This opens up a window into studying the relatively inaccessible yet vitally important Auger processes. Finally, we note that similar device geometries with hBN-separated FLG gates are used widely for electrostatic gating of 2D semiconductors under optical excitation, and Auger-assisted gate photocurrent should be incorporated as an important factor in analyzing the performance of these devices.

# Methods

### Device fabrication

A detailed description of the fabrication process can be found in the Supplementary Information (SI) §1 and §2. In brief, this was accomplished in three stages. In the first, individual flakes were obtained by mechanical exfoliation and identified under an optical microscope. Second, van der Waals assembly was undertaken with the aid of polycarbonate films stretched over viscoelastic stamps. For each device, a total of 8 or 9 nano-flakes were stacked according to the device geometry shown in SI §1. Finally, metal electrodes consisting of 10/50 nm vanadium/gold were defined by electron beam lithography and electron beam physical vapor deposition (EBPVD).

### Photoluminescence, photocurrent spectroscopy and charge modulation spectroscopy

All measurements presented in the main text were performed in a cold finger cryostat at a temperature of 5 K. For photoluminescence, a 660-nm beam from a pulsed supercontinuum laser was used, with the average power kept at 10 μW. $V_T$ was varied from -4 V to +4 V; both CT1 and CT2 were grounded while BG1 and BG2 were disconnected. The emission was collected in reflection geometry and spectrally resolved with a CCD-mounted spectrometer. Photocurrent and charge modulation absorption spectroscopy were performed concurrently with the setup shown in the SI §4. In brief, an ac modulation voltage is added to $V_T$ and the probe laser wavelength was scanned from 500 nm to 760 nm. The dc component of $I_C$ is measured with a current preamplifier and the ac component of the probe laser was detected with a Si photodiode connected to a lock-in amplifier. Both supercontinuum and tunable cw sources were used in photocurrent and optical absorption measurements, but no difference was observed in the results. Detailed post-processing steps and analysis of the optical absorption spectra are given in the SI §5.

# References

1. Splendiani, A. *et al.* Emerging Photoluminescence in Monolayer $MoS_2$. *Nano Lett.* **10,** 1271–1275 (2010).
2. Mak, K. F., Lee, C., Hone, J., Shan, J. & Heinz, T. F. Atomically Thin $MoS_2$: A New Direct-Gap Semiconductor. *Phys. Rev. Lett.* **105,** 136805 (2010).
3. Shi, H. *et al.* Exciton Dynamics in Suspended Monolayer and Few-Layer $MoS_2$ 2D Crystals. *ACS Nano* **7,** 1072–1080 (2013).
4. Kozawa, D. *et al.* Photocarrier relaxation pathway in two-dimensional semiconducting transition metal dichalcogenides. *Nat. Commun.* **5,** 4543 (2014).
5. Beattie, A. R., Landsberg, P. T. & Frohlich, H. Auger effect in semiconductors. *Proc. R. Soc. London. Ser. A. Math. Phys. Sci.* **249,** 16–29 (1959).
6. Konabe, S. & Okada, S. Effect of Coulomb interactions on optical properties of monolayer

transition-metal dichalcogenides. *Phys. Rev. B* **90,** 155304 (2014).
7. Mouri, S. *et al.* Nonlinear photoluminescence in atomically thin layered $WSe_2$ arising from diffusion-assisted exciton-exciton annihilation. *Phys. Rev. B* **90,** 155449 (2014).
8. Kumar, N. *et al.* Exciton-exciton annihilation in $MoSe_2$ monolayers. *Phys. Rev. B* **89,** 125427 (2014).
9. Sun, D. *et al.* Observation of Rapid Exciton–Exciton Annihilation in Monolayer Molybdenum Disulfide. *Nano Lett.* **14,** 5625–5629 (2014).
10. Javaux, C. *et al.* Thermal activation of non-radiative Auger recombination in charged colloidal nanocrystals. *Nat. Nanotechnol.* **8,** 206–212 (2013).
11. Cunningham, P. D., McCreary, K. M. & Jonker, B. T. Auger Recombination in Chemical Vapor Deposition-Grown Monolayer $WS_2$. *J. Phys. Chem. Lett.* **7,** 5242–5246 (2016).
12. Robert, C. *et al.* Exciton radiative lifetime in transition metal dichalcogenide monolayers. *Phys. Rev. B* **93,** 205423 (2016).
13. Wang, H., Zhang, C. & Rana, F. Ultrafast Dynamics of Defect-Assisted Electron–Hole Recombination in Monolayer $MoS_2$. *Nano Lett.* **15,** 339–345 (2015).
14. Cunningham, P. D. *et al.* Charge Trapping and Exciton Dynamics in Large-Area CVD Grown $MoS_2$. *J. Phys. Chem. C* **120,** 5819–5826 (2016).
15. Danovich, M., Zólyomi, V., Fal'ko, V. I. & Aleiner, I. L. Auger recombination of dark excitons in $WS_2$ and $WSe_2$ monolayers. *2D Mater.* **3,** 35011 (2016).
16. Malic, E. *et al.* Dark excitons in transition metal dichalcogenides. *Phys. Rev. Mater.* **2,** 14002 (2018).
17. Chandni, U., Watanabe, K., Taniguchi, T. & Eisenstein, J. P. Evidence for Defect-Mediated Tunneling in Hexagonal Boron Nitride-Based Junctions. *Nano Lett.* **15,** 7329–7333 (2015).
18. Choi, K. K., Levine, B. F., Bethea, C. G., Walker, J. & Malik, R. J. Negative differential photoconductance in an alternately doped multiple quantum well structure. *Appl. Phys. Lett.* **52,** 1979–1981 (1988).
19. Wang, Z., Zhao, L., Mak, K. F. & Shan, J. Probing the Spin-Polarized Electronic Band Structure in Monolayer Transition Metal Dichalcogenides by Optical Spectroscopy. *Nano Lett.* **17,** 740–746 (2017).
20. Nguyen, P. V. *et al.* Visualizing electrostatic gating effects in two-dimensional heterostructures. *arXiv:1904.07301* (2019).
21. Wilson, N. R. *et al.* Determination of band offsets, hybridization, and exciton binding in 2D semiconductor heterostructures. *Sci. Adv.* **3,** e1601832 (2017).
22. Britnell, L. *et al.* Field-Effect Tunneling Transistor Based on Vertical Graphene Heterostructures. *Science* **335,** 947–950 (2012).
23. Liu, G.-B., Xiao, D., Yao, Y., Xu, X. & Yao, W. Electronic structures and theoretical modelling of two-dimensional group-VIB transition metal dichalcogenides. *Chem. Soc. Rev.* **44,** 2643–2663 (2015).
24. Nakamura, A. & Morigaki, K. Photoconductivity Associated with Auger Recombination of Excitons Bound to Neutral Donors in Te-Doped Gallium Phosphide. *J. Phys. Soc. Japan* **34,** 672–676 (1973).
25. Zhang, C. *et al.* Probing Critical Point Energies of Transition Metal Dichalcogenides: Surprising Indirect Gap of Single Layer $WSe_2$. *Nano Lett.* **15,** 6494–6500 (2015).
26. Cassabois, G., Valvin, P. & Gil, B. Hexagonal boron nitride is an indirect bandgap semiconductor. *Nat. Photonics* **10,** 262–266 (2016).
27. Barrows, C. J. *et al.* Electrical Detection of Quantum Dot Hot Electrons Generated via a $Mn^{2+}$-Enhanced Auger Process. *J. Phys. Chem. Lett.* **8,** 126–130 (2017).
28. Tisdale, W. A. *et al.* Hot-electron transfer from semiconductor nanocrystals. *Science* **328,** 1543–7 (2010).
29. Pandey, A. & Guyot-Sionnest, P. Hot Electron Extraction From Colloidal Quantum Dots. *J. Phys. Chem. Lett.* **1,** 45–47 (2010).

**Acknowledgements**: This work was mainly supported by AFOSR FA9550-18-1-0104 and FA9550-18-1-0046, and partially supported by NSF EFRI (Grant No. 1741656). The initial measurement was supported by DoE BES (DE-SC0018171) through JRS. WY and HY were supported by the Croucher Foundation (Croucher Innovation Award), the RGC of Hong Kong (HKU17305914P), and the HKU ORA. DM and JY were supported by DoE BES, Materials Sciences and Engineering Division. K.W. and T.T. acknowledge support from the Elemental Strategy Initiative conducted by the MEXT, Japan and the CREST (JPMJCR15F3), JST. XX acknowledges support from the State of Washington through Clean Energy Institute and from the Boeing Distinguished Professorship in Physics.

**Author contributions**: XX, DHC, JRS, and CMC conceived the experiments. CMC fabricated the devices (with the assistance from PR), performed the experiments and analyzed the data, supervised by XX and DHC. WY and HY proposed the hole-Auger scattering mechanism. JRS conducted preliminary studies on a different sample, fabricated by JF. JY and DGM characterized and provided $WSe_2$ crystals. TT and KW provided hBN crystals. CMC, XX, DHC, and WY wrote the manuscript. All authors discussed the results.

**Author Information:** The authors declare no competing financial interests. Correspondence and requests for materials should be addressed to xuxd@uw.edu and cobden@uw.edu.

**Data Availability:** The data that support the findings of this study are available from the corresponding author upon reasonable request.

# Supplementary Information for
# Monolayer Semiconductor Auger Detector

Colin M. Chow[1], Hongyi Yu[2,3], John R. Schaibley[1,4], Pasqual Rivera[1], Joseph Finney[1], Jiaqiang Yan[5], David G. Mandrus[5,6], Takashi Taniguchi[7], Kenji Watanabe[7], Wang Yao[3], David H. Cobden[1]*, Xiaodong Xu[1,8]*

[1]Department of Physics, University of Washington, Seattle, Washington 98195, USA.
[2]Guangdong Provincial Key Laboratory of Quantum Metrology and Sensing and School of Physics and Astronomy, Sun Yat-Sen University (Zhuhai Campus), Zhuhai 519082, China.
[3]Dept. of Physics & Centre of Theoretical and Computational Physics, University of Hong Kong, Hong Kong, China.
[4]Department of Physics, University of Arizona, Tucson, Arizona 85721, USA.
[5]Materials Science and Technology Division, Oak Ridge National Laboratory, Oak Ridge, Tennessee 37831, USA.
[6]Department of Materials Science and Engineering, University of Tennessee, Knoxville, Tennessee 37996, USA.
[7]National Institute for Materials Science, Tsukuba, Ibaraki 305-0044, Japan.
[8]Department of Materials Science and Engineering, University of Washington, Seattle, Washington 98195, USA
*email: cobden@uw.edu, xuxd@uw.edu

**This supplementary document contains**

## §1. Fabrication steps

First, we search under microscope for suitable $WSe_2$ monolayers, few-layer graphene (FLG) and thin hBN flakes after mechanical exfoliation of bulk crystals onto $SiO_2$ substrates. For hBN, selection is primarily based on the area size (> 25 × 25 μm$^2$) and thickness (around 10 nm); while for FLG, the geometrical shape (ribbons are preferred), as well as the thickness (4 to 8 nm). Next, their surface quality and exact thickness are determined with an atomic force microscope (AFM). Only atomically-flat flakes are used.

Next, we use a layer-by-layer van der Waals stacking method[1] to assemble the device using a setup shown in Fig. S1a: the topmost flake is picked up by a polycarbonate (PC) film stretched over a viscoelastic polydimethylsiloxane (PDMS) dome, with the assistance of substrate heating (upto 120 °C). This process is repeated for all other flakes and the assembled stack is deposited onto the substrate by melting down the PC film. The residual PC is then removed by immersion in chloroform. To minimize the risk of fabrication errors, we split this stage into two parts as shown in Fig. S1b. First, FLG bottom gates BG1 and BG2, together with the lower hBN barrier, are transferred onto $SiO_2$ substrate, followed by thermal annealing at 400 °C for 30 minutes. The rest of the device is then deposited onto the pre-fabricated bottom gates and annealed at 200 °C for an additional 30 minutes. In addition, after every annealing step, we used AFM to assess the surface quality, and to cross check layer thicknesses.

Finally, electrical connections to the FLG gates and contacts are made by electron beam lithography and EBPVD of 5/50-nm-thick vanadium/gold composite. For the device discussed in the main text, the thickness of the upper and lower hBN is 7.6 nm and 10.3 nm, respectively, while that of BG1, BG2, CT1, CT2 and T is 4.3 nm, 8.3 nm, 7.3 nm, 4.3 nm and 6.0 nm, respectively.

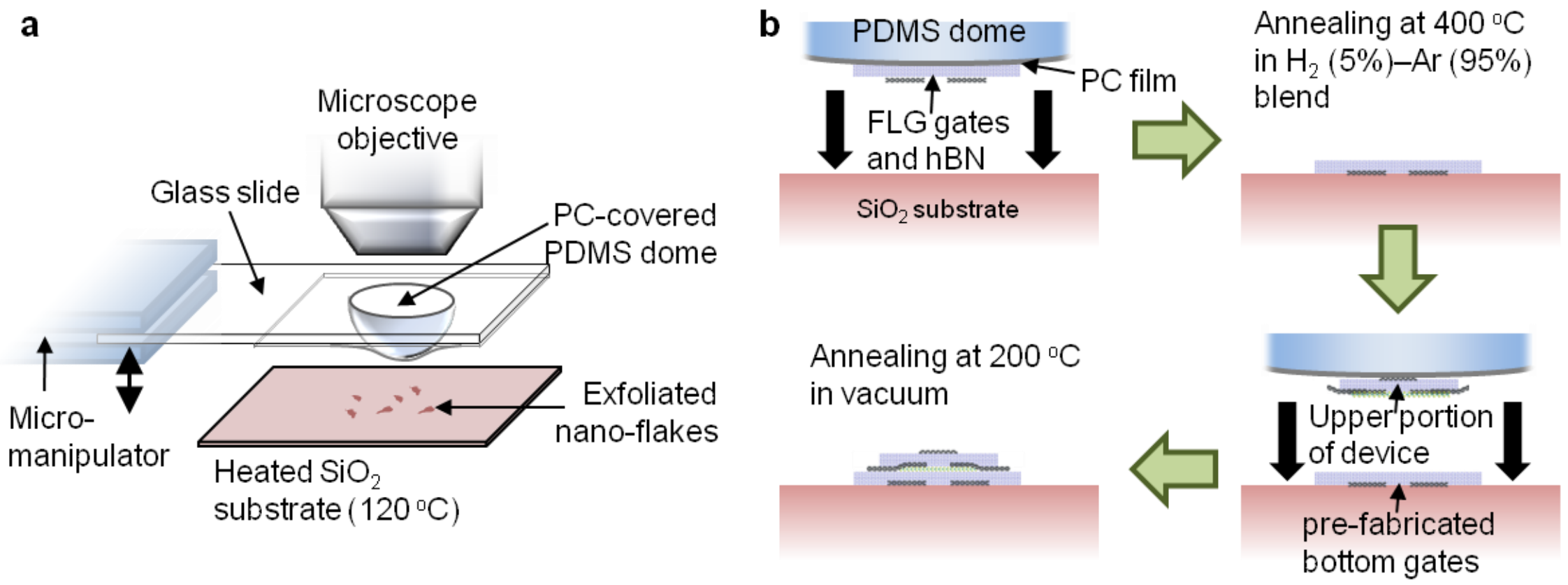

**Fig. S1 | Van der Waals stacking of nano-flakes to form devices. a,** Schematic illustration of the setup. **b,** Fabrication sequence consisting of two stack-and-anneal cycles.

## §2. Device characterizations

All experiments discussed in this work are carried out at a temperature of 5 K. For current measurements, a preamplifier (Ithaco DL1211) is used, along with a multifunction DAQ device (NI USB-6259) for input voltage control and data recording. Optical measurements are made in reflection geometry with an objective mounted on a piezo-actuated micrometer stage assembly. The laser beam is spatially filtered by a short single mode optical fibre, resulting in a focal spot size of about 1 μm in diameter.

For the initial device characterization, we measured the I-V characteristics of all electrode pair combinations. The result of that between T and CT1 is shown in Fig. 1d in the main text. Others are plotted in Figs. S2a and b. Nearly similar I-V characteristics are observed between electrode pairs T–CT1, T–CT2, BG1–CT1 and BG2–CT2. Results from T–BG1 and T–BG2 show the expected increase in the onset voltage bias of tunneling current due to effectively thicker hBN barrier. For all other combinations, negligible current is detected within ±8 V bias range. To avoid damaging the device, we operate in the low current plateau between tunneling onsets, *i.e.*, ± 4 V.

Conductance of the $WSe_2$ channel with respect to top gate/electrode bias, $V_T$, is shown in Fig. S2c. This measurement is taken with -3 V applied to both bottom gates ($V_{BG1} = V_{BG2}$) for $V_T < 0$ and +3 V for $V_T \geq 0$, while $V_{CT2}$ is fixed at 50 mV. Correspondingly, $V_T = \pm 3$ V is used to obtain the channel conductance as a function of $V_{BG1} = V_{BG2}$ shown in Fig. 1c in the main text. Fig. S2d plots the source-drain I-V curve measured at $V_T = V_{BG1} = V_{BG2} = 3$ V, which displays a nearly Ohmic behavior with a channel resistance of 82 MΩ.

In addition to electrical characterizations, we also measured $WSe_2$ photoluminescence spectra as a function of $V_T$. The result is shown in Fig. S2e. As indicated in the plot, several features of interest are: the A exciton ($X_A^0$) which dominates the spectra near zero $V_T$ and trion complexes that emerge with larger $V_T$ in both positive and negative directions. This result agrees well with other published works[2,3].

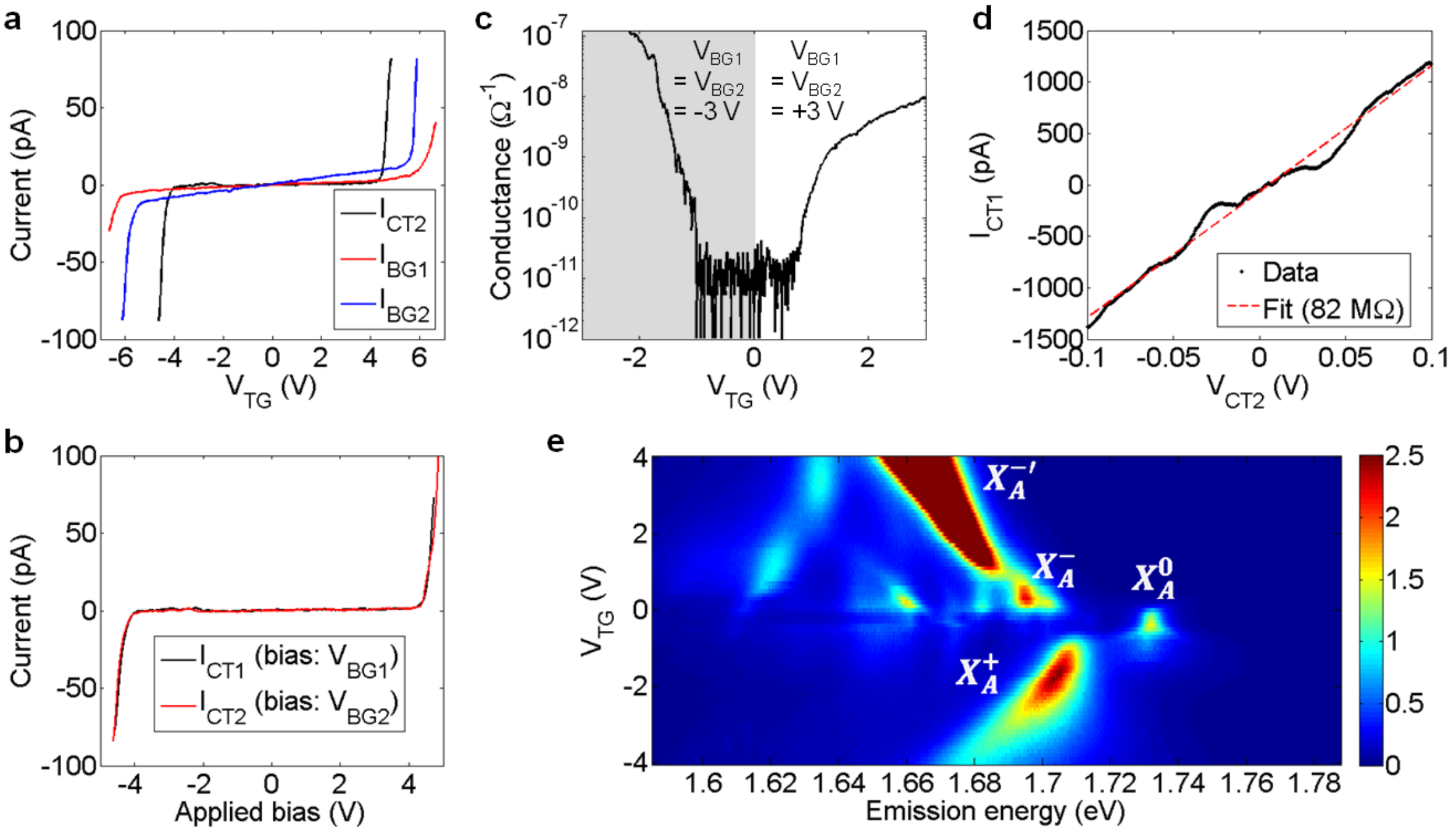

**Fig. S2 | Basic characterization results. a,** I-V curves between electrode pairs T–CT2, T–BG1 and TG–BG2. **b,** I-V curves between electrode pairs BG1–CT1 and BG2–CT2. **c,** $WSe_2$ channel conductance as a function of $V_T$. **d,** $I_{CT1}$ versus $V_{CT2}$ measured at $V_T = V_{BG1} = V_{BG2} = 3$ V. **e,** Intensity plot of $V_T$-dependent photoluminescence spectra. Colour bar: k-counts $s^{-1}$.

## §3. Operating condition for conventional photocurrent spectroscopy

The typical method for studying photoabsorption in transition metal dichalcogenides (TMDCs) is by measuring their reflectance spectra. Occasionally, photocurrent spectroscopy is also used. Although the latter requires a more complicated setup and a wavelength-tunable light source, it can be advantageous in certain cases. In the former technique, one typically measures the reflectance contrast between on- and off-sample photoexcitation. This can be challenging for probing photoactive layers embedded within multi-layered heterostructures like our device. The interposing layers may cause optical interference and, subsequently, a dispersive lineshape that complicates precise determination of the absorption resonance. This problem is mitigated in photocurrent spectroscopy where the signal is linearly proportional to the absorption at the excitation wavelength.

The versatility of our device in supporting multimodal studies is showcased in the main text. Here, we show how it can also be configured for conventional photocurrent spectroscopy with the bias configuration shown in Fig. S3a. Gate voltages of opposite polarity are applied to BG1 and BG2, while T is disconnected. This results in a p-i-n photodiode (Fig. S3b). As in the Auger photocurrent spectroscopy, a laser beam with tunable wavelength illuminates a small area (~ 1 $\mu m^2$) of the $WSe_2$ channel and current at one of the contacts is recorded. For the result shown in Fig. S3c, photocurrent from CT1 is recorded while CT2 is grounded, *i.e.*, zero-bias operation. When CT2 is probed instead, we obtained similar result but with the current flowing in the opposite direction, consistent with the band diagram shown in Fig. S3b. The spectrum shows a sharp resonance peak at 1.73 eV originating from the neutral exciton, $X_A^0$. At about 30 meV below, trion ($X_A^-$) is also visible as the shoulder of $X_A^0$. In addition, a small peak is observed at 130 meV above $X_A^0$. This has been attributed to either 2s excitonic Rydberg state[4–6], or exciton-phonon complex[7,8].

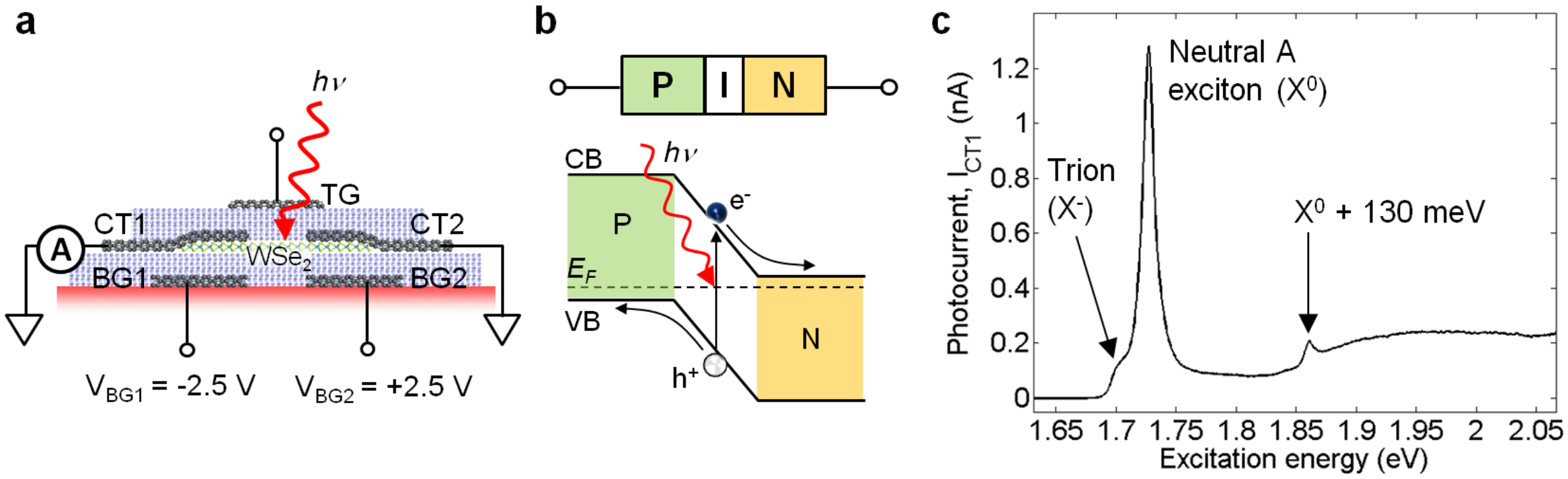

**Fig. S3 | Conventional photocurrent spectroscopy. a,** Device operating condition. **b,** Corresponding p-i-n photodiode (top) and band diagram (bottom). CB: conduction band; VB: valence band; $E_F$: Fermi level. **c,** Photocurrent spectrum measured with the configuration in **a**.

## §4. Experimental setup for concurrent Auger photocurrent and charge modulation spectroscopy

The Auger photocurrent and absorption intensity plots shown in Figs. 2a and 2b in the main text are produced from multimodal experiments performed simultaneously with the setup shown in Fig. S4a. A square wave of 0.1 $V_{\text{peak-to-peak}}$ at 1 kHz is added to $V_T$. The laser wavelength is varied from 500 nm to 760 nm with 1-nm step size. For the photocurrent measurement, dc current from CT1 is measured; while for the charge-modulation (CM) absorption, the ac component of the reflected laser intensity is recorded from the output of a lock-in amplifier.

Fig. S4b schematically illustrates the principle of CM spectroscopy. In a hypothetical scenario where the absorption resonance shows a bias-dependent frequency shift, the modulation of bias voltage results in oscillations of the resonance. When probed by a laser at a fixed wavelength, this modulation is imparted to the intensity of the reflected beam, whose amplitude is then measured at the modulation frequency. In the case of a square-wave modulation employed here, the signal level is proportional to the difference between the two frequency-shifted spectra, $R(V_+) - R(V_-)$, up to a constant phase factor.

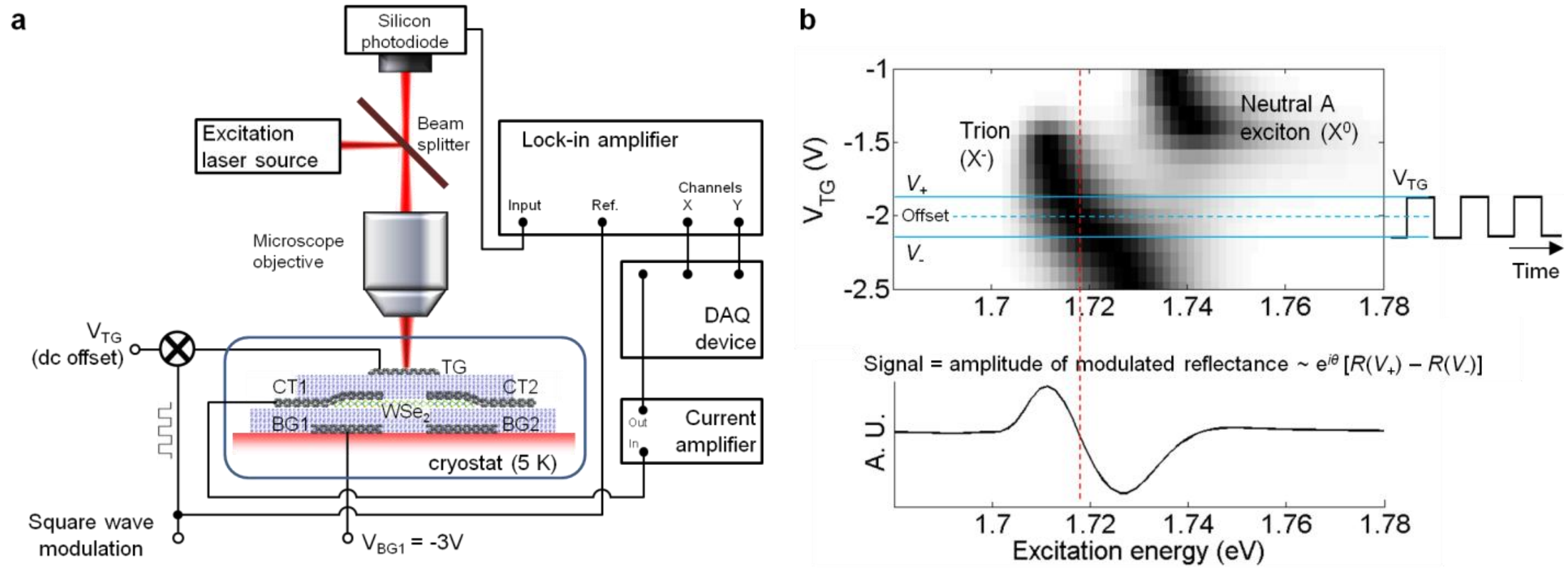

**Fig. S4 | Setup for concurrent Auger photocurrent and CM spectroscopy. a,** Experimental setup. **b,** Schematic illustration of the working principle of CM spectroscopy, using a hypothetical absorption intensity map.

## §5. CM spectroscopy data post-processing steps

The original CM spectroscopy data from the lock-in amplifier output is plotted in Fig. S5a. Here, each horizontal trace represents the difference between spectra from two bias points separated by $V_{\text{peak-to-peak}}$ of the modulation. To facilitate a meaningful comparison with the Auger photocurrent plot, post-processing is needed to reconstruct the actual bias-dependent absorption spectra. This is accomplished in 3 steps:

i. The data is interpolated in the direction of $V_T$ (vertical axis) such that the new step size of the bias offset (dc bias of $V_T$) equals $V_{\text{peak-to-peak}}$.
ii. An ansatz (a reasonable guess) is defined for the spectrum at the endpoint of the bias range, here chosen to be at $V_T$ = -4 V. The choice is completely arbitrary, but it is convenient to start from the side with sparse spectral features or negligible signal amplitude. For the result shown in Fig. 2b, the ansatz at $V_T$ = -4 V is plotted in Fig. S5b.
iii. Using the ansatz as the initial point, the original CM spectra are sequentially added to produce the actual absorption intensity map shown in Fig. 2b, reproduced here in Fig. S5c for convenience. Each horizontal trace here is the sum of all spectra below the corresponding $V_T$ in Fig. S5a.

When properly defined, the ansatz has very little effect on the overall appearance of the reconstructed absorption intensity plot, except in the first few bias steps. If improperly defined, however, it leaves vertical (*i.e.*, bias independent) streaks in the final result that can be easily recognized and corrected.

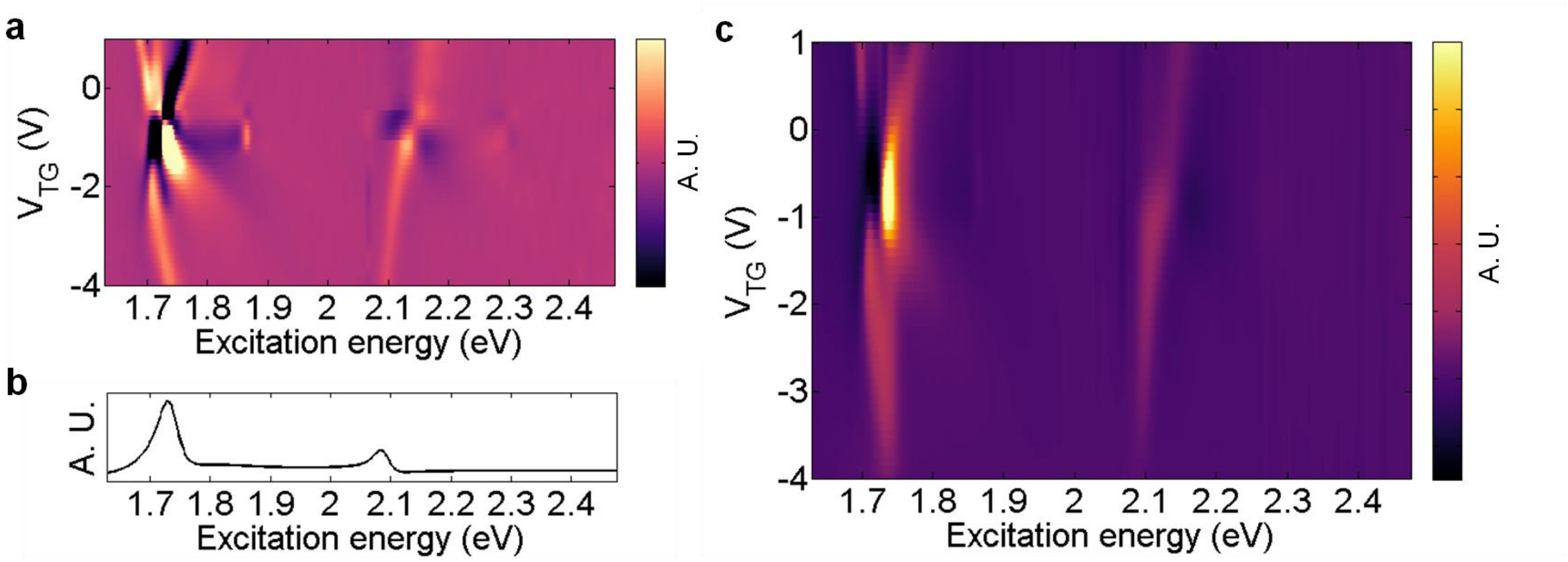

**Fig. S5 | CM spectra post-processing. a,** Raw data from the lock-in amplifier. **b,** Ansatz for initial point at $V_T$ = -4 V. **c,** Final result shown in Fig. 2b, reproduced here for convenience.

## §6. Data from a second device

The micrograph of our second device is shown in Fig. S6a. It has the same geometry as the first, but with upper and lower hBN thicknesses of 11.7 nm and 10.0 nm, respectively. In the p-doped regime (negative $V_T$), the Auger photocurrent spectra is qualitatively similar to those measured in the first device, with clearly distinguishable spectral features of trion, A and B excitons. (See Fig. S6b) In the n-doped regime, the spectra is mostly devoid of spectral feature, as in the first device. While positive photocurrent can still be seen, the region where it occurs is smaller than that in our first device. Despite this, we can still estimate the slope of its onset, around -11 $e^{-1}$, evaluated about the excitation energy of 2.45 eV. (See Fig. S6c) The steeper slope compared to that of the first device is due to the thicker top hBN layer used here. As will be shown in later sections, the steepness of the slope is proportional to the thickness of the hBN tunnel barrier.

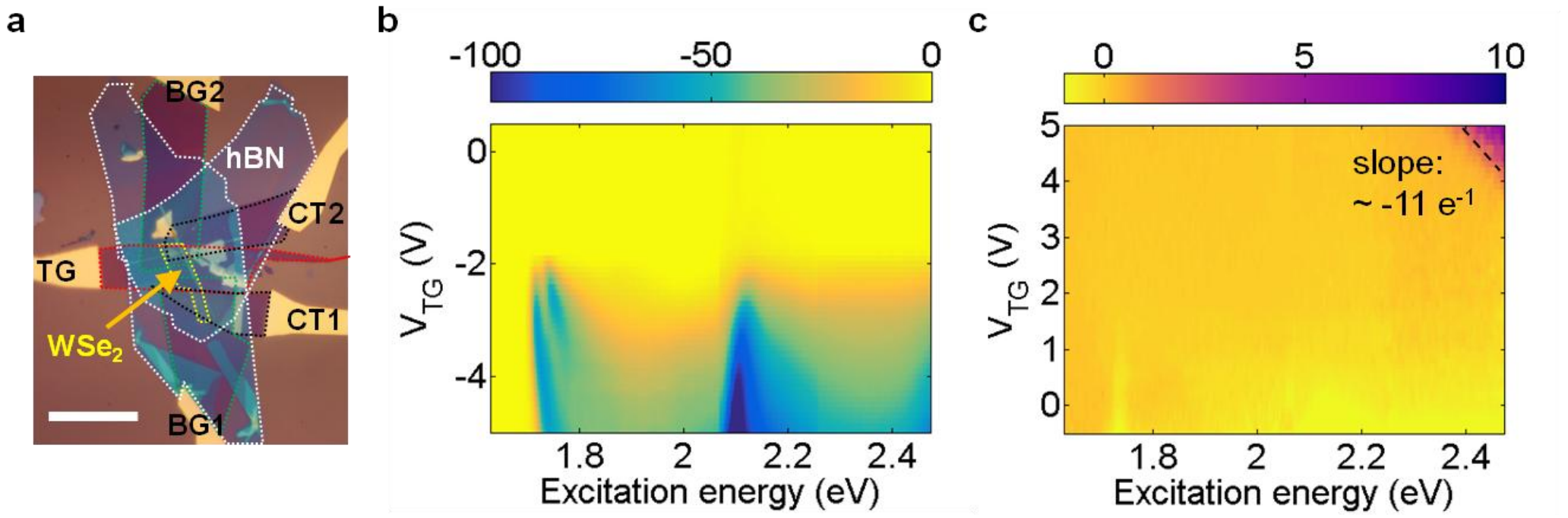

**Fig. S6 | Data from a second device. a,** Optical micrograph. Scale bar: 20 µm. **b,** $V_T$-dependent Auger photocurrent in p-doped regime. **c,** Photo-response in n-doped regime. Dashed line indicates the onset of positive photocurrent. Colour bar: pA/µW.

## §7. Estimation of exciton-hole Auger recombination rate

To produce a rough estimate of the exciton-hole Auger scattering rate, we consider the following rate equations:

$$\frac{dn_{X_A^+}}{dt} = R_{X_A^+} - n_{X_A^+}\left(\frac{1}{\tau_r}+\frac{1}{\tau_{Au}}+\frac{1}{\tau_{nr}}\right)$$
$$\frac{dn_{Au}}{dt} = \frac{n_{X_A^+}}{\tau_{Au}} - \frac{n_{Au}}{\tau_{th}} - J$$

where $n_{X_A^+}$ and $n_{Au}$ are densities of positive trion ($X_A^+$) and Auger hole, respectively. These two quantities are coupled through $1/\tau_{Au}$, the Auger rate to be estimated here. $R_{X_A^+}$ denotes the generation rate of $X_A^+$, which is a function of hole doping level and photo-excitation intensity. The radiative recombination of $X_A^+$ is characterized by the lifetime $\tau_r$, while any nonradiative recombination excluding the exciton-hole Auger scattering is characterized by $\tau_{nr}$. $J$ is the Auger photocurrent density to be inferred from our data. Here, we also include a variable $\tau_{th}$ to represent the loss of Auger holes in $WSe_2$ due to re-thermalization or other mechanisms.

Under photo-excitation with a cw source, we can assume that time-equilibrium is achieved so that the rate equations above equals zero. This leads to:

$$\frac{1}{\tau_{Au}} = \left(\frac{1}{\tau_r}+\frac{1}{\tau_{nr}}\right)\frac{J + n_{Au}/\tau_{th}}{R_{X_A^+} - (J + n_{Au}/\tau_{th})}$$

Note that with a simple rearrangement of the terms, the linear relationship between Auger photocurrent and photo-excitation intensity can be seen in:

$$J = \frac{R_{X_A^+}}{1 + \tau_{Au}(1/\tau_r + 1/\tau_{nr})} - \frac{n_{Au}}{\tau_{th}}$$

since both $R_{X_A^+}$ and $n_{Au}$ are linearly proportional to photo-excitation intensity under weak excitation.

Our estimate of the Auger rate depends on the term $n_{Au}/\tau_{th}$. This arises from the competition between the transmission of Auger hole across the hBN barrier and re-thermalization of the hot Auger hole. All these mechanisms are understudied at the moment. However, if we assume that transmission prevails over re-thermalization (i.e., $\tau_{th}$ is large), a lower bound of $1/\tau_{Au}$ can be estimated with

$$\frac{1}{\tau_{Au}} \geq \left(\frac{1}{\tau_r}+\frac{1}{\tau_{nr}}\right)\frac{J}{R_{X_A^+} - J}$$

The density of Auger photocurrent, $J$, can be obtained by assuming that it is generated only in the region illuminated by the excitation laser (~ 1 μm$^2$). From our Auger photocurrent data (Fig. 2a), $J \approx 5.3\times10^{16}$ e cm$^{-2}$ s$^{-1}$ μW$^{-1}$ at $V_T$ = - 4 V. To estimate $R_{X_A^+}$, we use the measured absorbance of 0.5 % at the $X_A^+$ resonance of 1.735 eV at $V_T$ = - 4 V (Fig. 2b), yielding $R_{X_A^+} = 1.8\times10^{18}$ cm$^{-2}$ s$^{-1}$ μW$^{-1}$. Therefore, we have $1/\tau_{Au} \geq 0.03(1/\tau_r + 1/\tau_{nr})$ . This value depends on the radiative and other nonradiative recombination times of $X_A^+$, of which experimental measurements are still lacking. Nonetheless, based on reported trion lifetime in the order of 1 ps in the literature[9–12], the value of $(1/\tau_r + 1/\tau_{nr})^{-1}$ is likely to fall within the same order of magnitude. In this case, $1/\tau_{Au}$ is estimated to be in the order of $10^{10}$ s$^{-1}$. The corresponding Auger coefficient, defined as $k = 1/n\tau_{Au}$, is then approximately 0.001 cm$^2$ s$^{-1}$. Here, $n = q^{-1}c_{hBN}V_T$ is the hole doping concentration inferred from the capacitance per unit area, $c_{hBN}$, of the device. Interestingly, our estimate of $k$ is of the same order of magnitude as that theoretically predicted for exciton-electron Auger scattering by Danovich et al.[13] It is also pointed out therein that this Auger coefficient is two orders of magnitude smaller than that of exciton-exciton Auger scattering which occurs at much higher excitation density.

As a comparison, Auger constants for bulk semiconductors typically fall in the range of $10^{-30}$ to $10^{-28}$ cm$^6$ s$^{-1}$ [14]. In the case of highly doped samples ($10^{16}$ – $10^{18}$ cm$^{-3}$, depending on the material), this corresponds to an upper limit of Auger rate in the order of $10^8$ s$^{-1}$. This is still two orders of magnitude less than the estimate of trion-hole Auger rate in $WSe_2$ monolayer as inferred from our measurement.

## §8. Onset of positive photocurrent based on photo-assisted tunneling

In this scenario, we assume that the change in Fermi level with respect to applied bias, $V_T$, is small such that the Fermi level remains near conduction band edge. According to Fig. 4d, an incident photon excites a Fermi electron to an excited state from which the electron can tunnel through the top hBN barrier. The applied bias tilts the conduction band edge of the hBN barrier, resulting in a triangular potential barrier whose effective width reduces with increasing $V_T$. If we can further treat the $WSe_2$ monolayer as a quantum well (QW), the photocurrent is then directly proportional to the escape rate of a photo-excited electron, which is given by the expression[15]:

$$R = \frac{1}{\tau} = A(d_{QW}, E_{ph})\exp\left[-\frac{8}{3}\frac{\sqrt{2m^*}(\Delta E)^{\frac{3}{2}}}{e\hbar F}\right] \quad (4)$$

where $\tau$ is the mean lifetime and $F$ the electric field across the hBN barrier which defines the band tilting, and is given by $V_{TG}/t_{hBN}$. Since we assumed that the electron is excited to a state with energy $E_{ph}$ above the conduction band edge, the effective barrier height $\Delta E$ becomes $E_{offset} - E_{ph}$. The exponential factor is derived from WKB method while the pre-exponential factor $A(d_{QW}, E_{ph})$, related to the impingement rate, depends on various parameters such as the initial electron energy level (assumed to be $E_{ph}$ here) and the exact QW geometry, and its choice remains a debated subject[16]. Here, using a simple kinetic energy model, we let $A(d_{QW}, E_{ph}) = t_{\mathrm{WSe_2}}{}^{-1}\sqrt{E_{ph}/2m^*}$, where $t_{\mathrm{WSe_2}}$ is the effective QW thickness of $WSe_2$. In doing so we assume that the electron wavefunction is largely confined to the $WSe_2$ layer due to the presence of van der Waals gaps. However, we also examined a few other forms of prefactors and found that they all have a negligible effect on our analysis because variations in tunneling rate is largely determined by the exponential factor.

Fig. S7 below is a relative magnitude plot of Eq. (4) with a top hBN thickness of 7.6 nm (first sample). We let $E_{offset}$ to be 3.0 eV and $t_{\mathrm{WSe_2}}$ an arbitrary value since it contributes to only a constant scaling factor. Evidently, within the experimentally accessible region indicated by the red rectangle, the calculated tunneling rate resembles our data shown in Fig. 3a. Therefore, the onset curve of positive photocurrent is most likely determined by the change in tunneling rate due to bias-induced band tilting of hBN barrier.

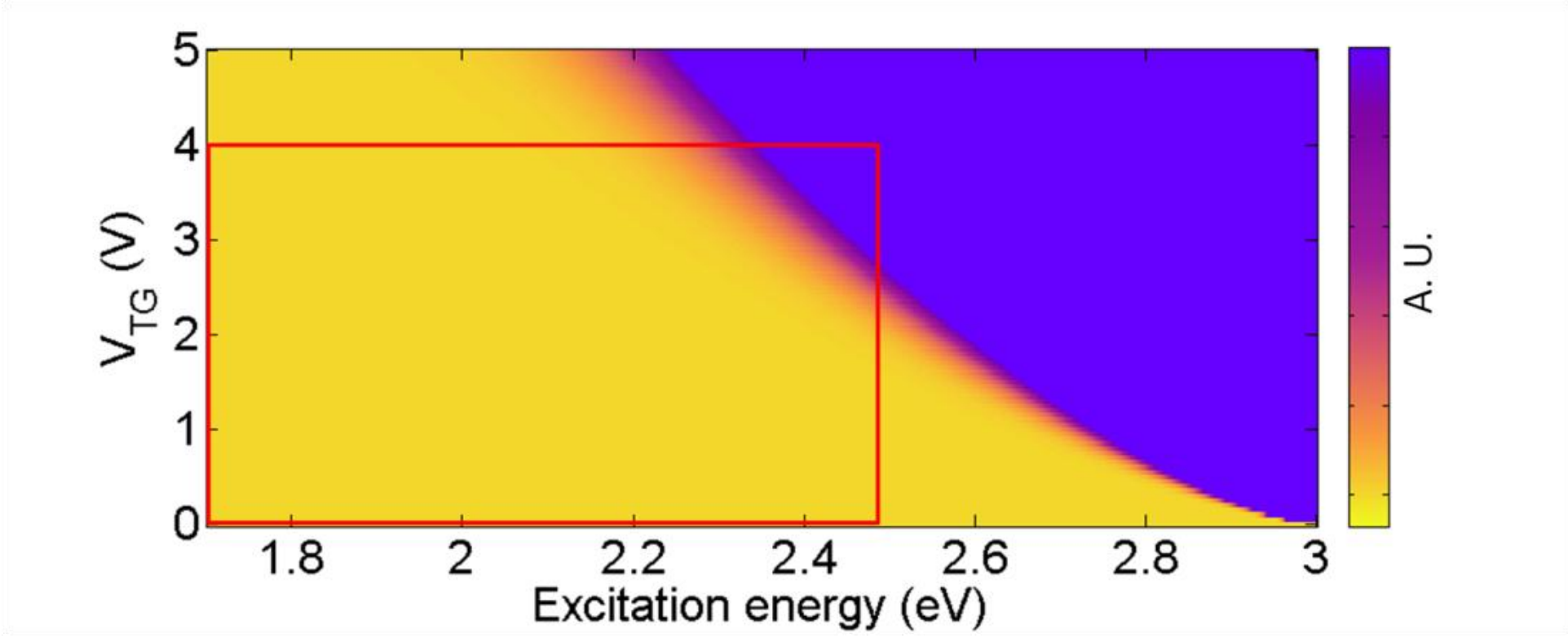

**Fig. S7 | Relative magnitude of tunneling rate as a function of $V_T$ and excitation photon energy.** Top hBN thickness of 7.6 nm is used, based on our first sample; while the $WSe_2$-hBN conduction band offset is assumed to be 3.0 eV. Experimentally accessible range is indicated by the red rectangle.

Although we assumed $E_{offset}$ to be 3.0 eV in our analysis above, we now show that we can instead estimate this value from the data with the aid of Eq. (4). First, solving for $V_T$, we have

$$V_T = \frac{-8\sqrt{2m^*}t_{hBN}\left(E_{offset} - E_{ph}\right)^{\frac{3}{2}}}{3e\hbar \ln\left(Rt_{\mathrm{WSe_2}}\sqrt{2m^*/E_{ph}}\right)}$$

Differentiate with respect to $E_{ph}$ gives

$$e\frac{dV_T}{dE_{ph}} = \frac{4\sqrt{2m^*\left(E_{offset} - E_{ph}\right)}t_{hBN}}{\hbar \ln\left(Rt_{\mathrm{WSe_2}}\sqrt{2m^*/E_{ph}}\right)}\left[1 - \frac{E_{offset} - E_{ph}}{3E_{ph}\ln\left(Rt_{\mathrm{WSe_2}}\sqrt{2m^*/E_{ph}}\right)}\right]$$

Note that the photocurrent onset follows the contour of a constant magnitude $R$, ignoring the proportionality constant. Now substitute $R$ in above equation with Eq. (4) and after some algebraic manipulations, we get

$$\left(E_{offset} - E_{ph}\right)^{\frac{3}{2}} + \frac{3V_T}{2\,dV_T/dE_{ph}}\sqrt{E_{offset} - E_{ph}} + \frac{3e\hbar V_T^2}{16t_{hBN}E_{ph}\sqrt{2m^*}\,dV_T/dE_{ph}} = 0$$

This is a depressed cubic equation of the form $x^3 + px + q = 0$ where $p >> q$, of which a physically meaningful, real-valued solution is given by the Cardano formula:

$$x = \sqrt[3]{-\frac{q}{2} + \sqrt{\frac{q^2}{4} + \frac{p^3}{27}}} + \sqrt[3]{-\frac{q}{2} - \sqrt{\frac{q^2}{4} + \frac{p^3}{27}}} \quad (5)$$

Here, $x = \sqrt{E_{offset} - E_{ph}}$, $p = \frac{3V_T}{2dV_T/dE_{ph}}$ and $q = \frac{3e\hbar V_T^2}{16t_{hBN}E_{ph}\sqrt{2m^*}dV_T/dE_{ph}}$. Referring to the data from our first device as shown in Fig. 3a, $e\frac{dV_T}{dE_{ph}}$ is estimated to be -7.4 about the point ($E_{ph}$, $V_T$) = (2.4 eV, 3.1 V). Substituting these into Eq. (5) above gives $E_{offset} = 3.03$ eV. Likewise, from Fig. S6c, with $e\frac{dV_T}{dE_{ph}} = -11$ about ($E_{ph}$, $V_T$) = (2.45 eV, 4.4 V), we have $E_{offset} = 3.06\,\text{eV}$. Thus, both of our samples yield an estimated $E_{offset}$ at around 3.0 eV, which is remarkably close to the expected value of 2.96 eV.

# References

1. Zomer, P. J., Guimarães, M. H. D., Brant, J. C., Tombros, N. & van Wees, B. J. Fast pick up technique for high quality heterostructures of bilayer graphene and hexagonal boron nitride. *Appl. Phys. Lett.* **105,** 13101 (2014).
2. Wang, Z., Zhao, L., Mak, K. F. & Shan, J. Probing the Spin-Polarized Electronic Band Structure in Monolayer Transition Metal Dichalcogenides by Optical Spectroscopy. *Nano Lett.* **17,** 740–746 (2017).
3. Jones, A. M. *et al.* Excitonic luminescence upconversion in a two-dimensional semiconductor. *Nat. Phys.* **12,** 323–327 (2016).
4. Manca, M. *et al.* Enabling valley selective exciton scattering in monolayer $WSe_2$ through upconversion. *Nat. Commun.* **8,** 14927 (2017).
5. Courtade, E. *et al.* Charged excitons in monolayer $WSe_2$: Experiment and theory. *Phys. Rev. B* **96,** 85302 (2017).
6. Stier, A. V. *et al.* Magnetooptics of Exciton Rydberg States in a Monolayer Semiconductor. *Phys. Rev. Lett.* **120,** 57405 (2018).
7. Jin, C. *et al.* Interlayer electron–phonon coupling in $WSe_2$/hBN heterostructures. *Nat. Phys.* **13,** 127–131 (2017).
8. Chow, C. M. *et al.* Unusual Exciton–Phonon Interactions at van der Waals Engineered Interfaces. *Nano Lett.* **17,** 1194–1199 (2017).
9. Yan, T., Ye, J., Qiao, X., Tan, P. & Zhang, X. Exciton valley dynamics in monolayer $WSe_2$ probed by the two-color ultrafast Kerr rotation. *Phys. Chem. Chem. Phys.* **19,** 3176–3181 (2017).
10. Yan, T., Qiao, X., Liu, X., Tan, P. & Zhang, X. Photoluminescence properties and exciton dynamics in monolayer $WSe_2$. *Appl. Phys. Lett.* **105,** 101901 (2014).
11. Wang, G. *et al.* Valley dynamics probed through charged and neutral exciton emission in monolayer $WSe_2$. *Phys. Rev. B* **90,** 75413 (2014).
12. Godde, T. *et al.* Exciton and trion dynamics in atomically thin $MoSe_2$ and $WSe_2$: Effect of localization. *Phys. Rev. B* **94,** 165301 (2016).
13. Danovich, M., Zólyomi, V., Fal'ko, V. I. & Aleiner, I. L. Auger recombination of dark excitons in $WS_2$ and $WSe_2$ monolayers. *2D Mater.* **3,** 35011 (2016).
14. Levinshteĭn, M. E., Rumyantsev, S. L. & Shur, M. *Handbook series on semiconductor parameters*. (World Scientific, 1996).
15. Shankar, R. *Principles of Quantum Mechanics*. (Springer Verlag, 2014).
16. Larkin, I. A., Ujevic, S. & Avrutin, E. A. Tunneling escape time from a semiconductor quantum well in an electric field. *J. Appl. Phys.* **106,** 113701 (2009).